 \documentclass[aps,pre,reprint]{revtex4-2}
\usepackage[latin1]{inputenc}
\usepackage{amsmath}
\usepackage{amsfonts}
\usepackage{amssymb}
\usepackage[dvipdfmx]{graphicx}
\usepackage{xcolor}
\usepackage{bm}
\usepackage[a4paper,top=2cm,bottom=2cm,left=2cm,right=2cm]{geometry}

\begin{document}
	
\author{Takahiro Sakaue}
\affiliation{Department of Physical Sciences, Aoyama Gakuin University,
5-10-1 Fuchinobe, Chuo-ku, Sagamihara, Japan}
\email{sakaue@phys.aoyamna.ac.jp}
\author{Enrico Carlon}
\affiliation{Soft Matter and Biophysics, KU Leuven, Celestijnenlaan 200D,
B-3001, Leuven, Belgium}

\title{Compounding formula approach to chromatin and active polymer dynamics}
	
\begin{abstract}
Active polymers are ubiquitous in nature, and often kicked by
persistent noises that break detailed balance. In order to capture
the out-of-equilibrium dynamics of such active polymers, we propose a
simple yet reliable analytical framework based on a {\it compounding
formula}.  Connecting polymeric dynamics to the isolated monomeric
behavior via the notion of {\it tension propagation}, the formula
allows us to clarify rich scaling scenarios alongside corresponding
intuitive physical pictures. We demonstrate distinctive transient and
steady-state scalings due to the non-Markovian nature of the active
noise. Aside from a paradigmatic example of an active Rouse polymer,
we expect the framework to be applicable to wide variety of spatially
extended systems including more general polymers (crumpled globule,
semiflexible polymers etc), fluctuation of growing interfaces, and an
array of particles in single-file configuration.
\end{abstract}

\maketitle

{\it Introduction} --
Chromatin, the complex of DNA and histone proteins found in living
cells, provides a paradigmatic example of an active polymeric system
whose dynamics have been intensively studied by experiments and
theoretical modeling. Chromatin exibits non-equilibrium dynamics
due to the presence of various ATP-driven molecular motors such
as loop extruders, RNA polymerases or histone remodeling enzymes
\cite{webe12,Javer_2013,Michieletto_2019,tort20,Zidovska_2021,garini24,forte26}.
Chromatin dynamics is usually analyzed
experimentally by tagging some specific sites
with fluorophores, whose positions are then tracked over time
\cite{Sinha2008,Garini_2015,lamp16,Dudko_2019,Maeshima_2020,Kimura_2022,gabr22,Keizer_2022,sala22,Gregor_2023}.
The mean-squared displacement (MSD) of the tagged locus shows an
anomalous dynamics, in many cases fitting the passive Rouse model
behavior $\sim \tau^{1/2}$, where $\tau$ is a time scale between two
observation times, but often showing complex crossovers~\cite{Kimura_2022}
and sometimes superdiffusion $\sim \tau^\alpha$ (with $\alpha > 1$)
which are likely signatures of underlying active processes dominating
over thermal noise~\cite{Zidovska_2021,Javer_2013}.  Much of our
knowledge about the active chromatin dynamics stems from the analysis
of the active Rouse model (see below), which can be solved exactly
\cite{vand15,vand17,Put_2019,osma17,eise17,Gompper_2020,Samanta_2016,goy24,polo25},
or numerical simulation of the related
models~\cite{Lowen_2014,Shin_2015,Hyeon_2018,Smrek_2020,Hiraiwa_2022,Onuchic_2024}.
Still, our current understanding lags behind the equilibrium
counterpart largely due to the lack of a clear-cut physical picture
behind the rich non-equilibrium dynamics.  In the literature, two
conflicting scaling predictions have been made based on the normal mode
analysis~\cite{vand15,vand17,Sakaue_2017,Put_2019,osma17,eise17,Gompper_2020}.
But the origin of the discrepancy as well as the underlying mechanism
behind each prediction remain to be known.  The aim of this paper is to
develop a scaling theory of the active Rouse model dynamics which will
allow us to obtain a deeper understanding of the physical origin of the
observed behaviors. In addition, the scaling theory can be extended to
cases which are not exactly solvable. The spirit is very much the same
as in the equilibrium Rouse model: from the analysis of such a model
one develops scaling insights that can be used to infer, for instance,
the effect of self-avoidance, which cannot be derived from exact
calculations~\cite{doi88,Panja_2010,Sakaue_2013,Amitai13,Saito_2015}.

In the Rouse model the time evolution of $z(n,t)$ a cartesian 
component of the $n^{th}$ monomer position is governed by the 
following Langevin equation
\begin{eqnarray}
\gamma \frac{\partial z(n,t)}{\partial t} = 
k \frac{\partial^2 z(n,t)}{\partial n^2} + f(n,t)
\label{eq:Rouse}
\end{eqnarray}
where $\gamma$ is the monomeric friction coefficient and $k$ is the spring
constant.  We use a continuous description with $n$ a real number which
assumes both positive and negative values, as we consider an infinitely
long polymer. Equation~(\ref{eq:Rouse}) represents the force balance
in the overdamped limit with $k \partial^2_n z(n,t)$ the elastic force
on the $n^{th}$ monomer due to the neighboring monomers and $f(n,t)$
all other forces, possibly including active ones.  We will consider
forces of stochastic origin $\langle f \rangle = 0$ and characterized
by the correlator
\begin{eqnarray}
    \langle f(n,t)  f(n',t')\rangle = A g(|t-t'|) \delta(n-n')
    \label{ff_corr}
\end{eqnarray}
with $A$ measuring the noise strength. The thermal (passive) noise limit
is $A=2 \gamma k_BT$ and $g(u)=\delta(u)$, but we will develop here an
analysis valid for a generic function $g(u)$, which could contain an
active and a passive component. Considering a long polymer and neglecting
end terms effects, we build up the general solution from the propagator
\begin{eqnarray}
G(n,t) = \left( \frac{\tau_0}{4 \pi t} \right)^{1/2} 
\exp{\left( -\frac{\tau_0 }{4 t}n^2 \right)}
\label{Green_func}
\end{eqnarray}
which solves \eqref{eq:Rouse} for $f=0$, with the monomeric time scale
defined as $\tau_0 = \gamma/k$. We recall that the normal mode approach
to solve the Rouse model \cite{doi88} typically applies to finite length
chains with $0 \leq n \leq N$. In the equilibrium Rouse model the tagged
monomer MSD shows a crossover from an anomalous ($\sim \tau^{1/2}$)
to a regular ($\sim \tau$) diffusive behavior at the Rouse time scale
$\tau_R \simeq \tau_0 N^2$.  In our case ($ N \gg 1$), we assume that the
crossover to ordinary center of mass diffusion regime for times beyond the
Rouse time $\tau_R$ exceeds any time scale of relevance for the system.

\begin{figure}[t]
    \centering
    \includegraphics[width=\linewidth]{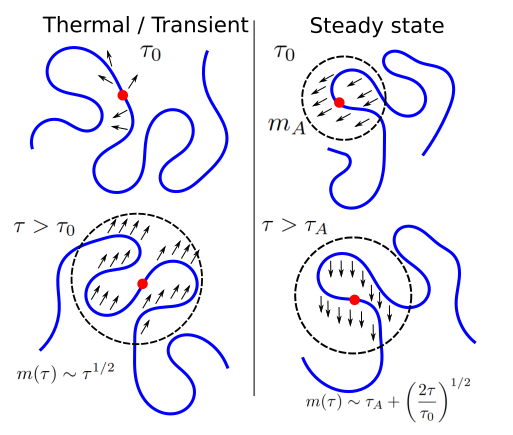}
    \caption{The compounding formula \eqref{CF2} connects the 
    MSD of a tagged monomer of the chain (in red) to that of an 
    isolated  monomer with an effective friction $m(\tau) \gamma$, 
    with $m(\tau)$ the number of monomers which are dynamically 
    connected to the tagged one. In active polymers $m(\tau)$, 
    represented by the monomers within the dashed circle, has a 
    richer scaling behavior as compared to equilibrium counterpart 
    (see text).}
    \label{fig01}
\end{figure}
{\it Tension propagation and compounding formula} --
The anomalous texbook MSD scaling $\sim \tau^{1/2}$ of a monomer in thermal 
equilibrium, as obtained from the solution of \eqref{eq:Rouse}, is
rationalized as being the consequence of the surrounding monomers along 
the polymer backbone. With increasing time scale, the tagged monomer drags
a growing number of segments on the chain. The contour length $m(\tau)$ of 
the dynamically correlated segments scales as $m(\tau) \sim \tau^{1/2}$ 
\cite{Rubinstein_book}, due to the underlying diffusive scaling of tension 
propagation \eqref{Green_func}. This results in an effective friction that 
increases as $m(\tau)$, thereby reducing the instantaneous mobility and 
slowing the diffusive motion by a factor $1/m(\tau) \sim \tau^{-1/2}$. By 
combining the bare diffusive behavior, characterized by an MSD $\sim \tau$, 
with this time scale-dependent drag, one obtains the subdiffusive equilibrium 
Rouse scaling $\text{MSD} \sim \tau^{1/2}$. This physical picture can be
formalized through the following compounding formula  
\begin{eqnarray}
     \langle \Delta z^2(n,\tau) \rangle \simeq
 \frac{\langle \Delta z_i^2(\tau) \rangle }{m(\tau)}
  \label{CF2}
 \end{eqnarray}
where the symbol $\simeq$ denotes a quasi-equality, i.e. it neglects
multiplicative factor of order unity. The previous formula expresses
the MSD of the $n^{th}$ monomer $\langle \Delta z^2(n,\tau) \rangle$
\footnote{For simplicity we focus on a single cartesian component and
refer to $\langle \Delta z^2\rangle$ as the MSD. The full MSD for a 3D
dynamics would comprise the contributions of the $x$ and $y$ components
as well.} over a time scale $\tau$ as the ratio of the MSD of an isolated
monomer $\langle \Delta z_i^2(\tau) \rangle$ with a factor $m(\tau)$ that
accounts for the number of monomers which are ``dynamically correlated"
to the $n^{th}$ monomer over a time scale $\tau$, Fig.~\ref{fig01}.
The compounding formula \eqref{CF2} was implicitly employed in the
analysis of tagged-monomer equilibrium dynamics \cite{Rubinstein_book},
but its usefulness and predictive power in more  complex polymer
architectures and non-equilibrium situations is still unexplored. This
is the scope of this letter.

The identification of the relevant characteristic length
scales is fundamental for elucidating the behavior of complex
systems. These concepts underpin the blob model, which
has been successfully employed to describe polymers under
tension~\cite{Pincus_1976,deGennes_book,Rubinstein_book}.
Blobs of various other types, such as those originating
from electrostatic interactions~\cite{deGennes_1976},
torsional constraints~\cite{Carlon_2015}, or topological
effects~\cite{Grosberg_2000, Sakaue_Wada_2010}, as well as
those associated with concentration fluctuations in semidilute
solutions~\cite{desCloizeaux1990,deGennes_book,Rubinstein_book},
have likewise been successfully employed.  The present work builds on
these concepts to address the dynamics of active polymers. The quantity
$m(\tau)$ can be interpreted as a dynamical blob comprising correlated
monomers. The application of Eq.~\eqref{CF2}, as we shall discuss, leads
to a richer phenomenology that has no analogue in thermal equilibrium.

While our focus here is on \eqref{eq:Rouse}, we show in the companion
paper \cite{long_paper} that the compounding formula works also for a
broader class of non-Gaussian models (referred to as $\beta$-models in the
polymer literature \cite{Amitai13}). We demonstrate that the rich spectrum
of dynamical scaling behaviors observed in active Rouse models can be
thus rationalized from the behavior of the terms on the right hand side
of \eqref{CF2}, which can be understood and analyzed more transparently.

\begin{figure}[t!]
    \centering
    \includegraphics[width=0.95\linewidth]{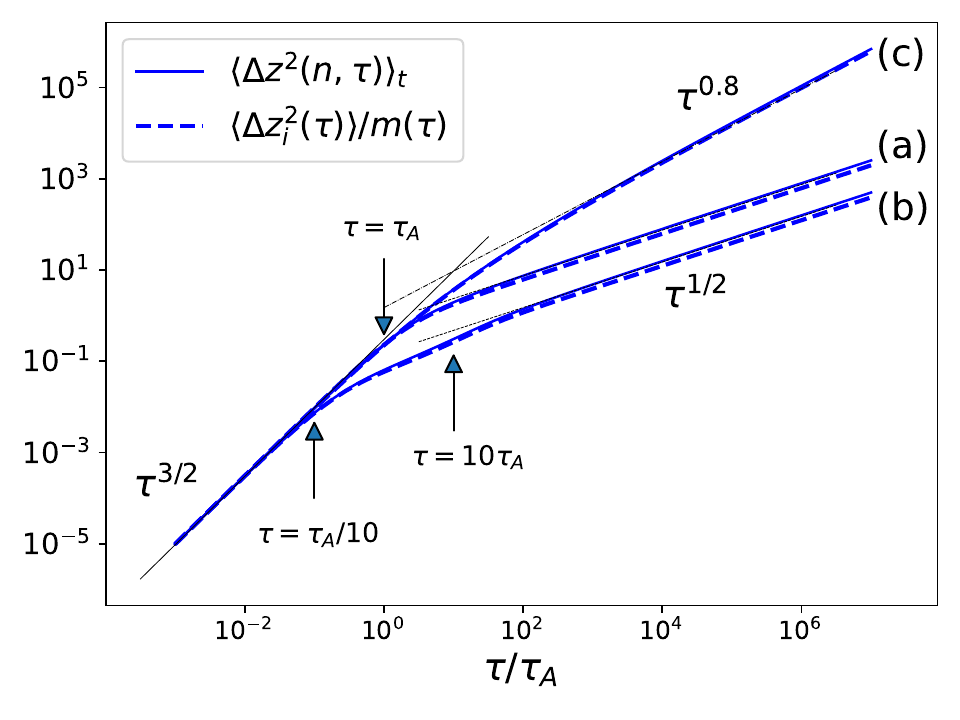}
    \caption{Test of the validity of the compounding formula
    \eqref{CF2} for the transient case and three different noises:
    (a) $g(u) = e^{-u/\tau_A}$, (b) $g(u) = [100 \, e^{-10u/\tau_A} +
    e^{-u/(10\tau_A)}]/101$ and (c) $g(u) = 1/[1+(u/\tau_A)^{0.7}]$. These
    are all normalized so that $g(0)=1$. The solid lines show the exact
    result on the tagged monomer MSD of a connected polymer. The dashed
    lines are plots of the ratio between Eqs.~\eqref{isolated_monomer}
    and \eqref{m_tau} (top).}
    \label{fig:compound}
\end{figure}

\begin{figure}[t!]
    \centering
    \includegraphics[width=0.95\linewidth]{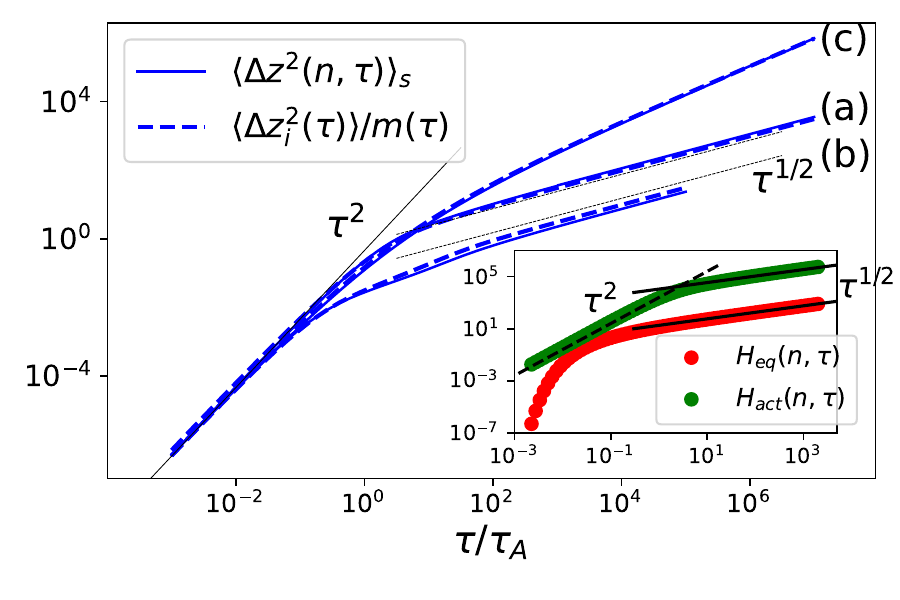}
    \caption{Test of the compounding formula for the steady state
    case for the same three noises (a,b,c) as for the transient
    case of Fig.~\ref{fig:compound}.  Inset: plot of $H(\tau,n)$
    vs. $\tau/\tau_A$ and fixed $n$ for thermal dynamics (red circles)
    and active dynamics in steady state with exponentially correlated
    noise (green circles). The latter shows a short time power law
    ($\sim \tau^2$) behavior.}
    \label{fig:compound-ss}
\end{figure}

{\it{Scaling predictions --}}
The isolated monomer MSD can be obtained from the solution of \eqref{eq:Rouse} 
with $k=0$ (thereby eliminating the chain connectivity) as
\begin{eqnarray}
    \langle \Delta z_i^2 (\tau)\rangle &=& \frac{2A}{\gamma^2}  
    \int_{0}^{\tau} du \ (\tau-u) \  g(u)
\label{isolated_monomer}
\end{eqnarray}
For an active Ornstein-Uhlenbeck (AOU) noise with $g(u) = e^{-u/\tau_A}$ 
in Eq.~\eqref{ff_corr}, we get
\begin{eqnarray}
    \langle \Delta z_i^2 (\tau)\rangle & \simeq & \frac{A}{\gamma^2}
\left\{
\begin{array}{cll} 
   \tau^2   &  &( \tau \ll \tau_A) \\   
   \tau_A \tau    & & ( \tau \gg \tau_A)
\end{array}
\right.
\label{msd-i}
\end{eqnarray}
where we consider $\tau_A \gg \tau_0$, otherwise the problem reduces
to an effective white noise correlator. The AOU noise preserves its
magnitude ($\sqrt{A}$) and direction over the time scale $\tau_A$,
and randomizes subsequently. Although the analysis can be extended to
the case with more slowly decaying noise (see Figs~\ref{fig:compound}
and~\ref{fig:compound-ss}), we employ such a simple form for clarity
unless otherwise noted.

We now consider $m(\tau)$, which arises from mechanisms
of tension propagation, i.e. the spreading of mechanical
forces through the polymer chain. Tension propagation
is central in the analysis of dynamical processes in polymers such as
translocation~\cite{Sakaue_2007,Sakaue_2010,Grosberg_2011,Ikonen_2013,Slater_2018,Saito_2012,Wanunu_2017,Sarabadani_2020,Micheletti_2017},
stretching~\cite{saka12,Grosberg_2012}, folding and
relaxation~\cite{fred14,Walter_2014,Vanderzande_2017}, where the factor
$m(\tau) \gamma$, the time scale-dependent friction is identified
as a source of the anomalous dynamics.  A fundamental difference
between thermal and active cases is that $m(\tau)$ crucially depends
on the protocol employed \cite{long_paper}.  We consider two different
situations.

In the {\it transient} protocol, the polymer remains in its ground
state until the noise is switched on at time $s$ and the MSD is
calculated from the differences in monomer positions between time $s$
and $t$. Mathematically, this is realized by writing the noise term in
Eq.~\eqref{eq:Rouse} as $\theta(t-s) f(t)$, where $\theta(t)$ is the step
function.  In the {\it steady state} protocol, the noise is switched on
in a distant past, at a time $-T_\infty$, with $T_\infty$ much larger
than any other characteristic relaxation times of the system. Under
these assumptions, the system at time $s$ is in a steady state. These
two protocols can be defined for a system driven by any noise, and we
will below elaborate the transient and steady state dynamics of the
thermal as well as the active Rouse models~\footnote{In more realistic
transient protocol for active polymer, the polymer is assumed to be
settled in thermal equilibrium until the active noise is turned on. Such
a thermal contribution to MSD can be calculated separately from the
active contribution~\cite{long_paper}.}.

The propagator \eqref{Green_func} implies that any local perturbation
applied at monomer $n$ at time $t=0$ grows with a variance linear in
time $\sigma^2 (t) = 2 t/\tau_0$. The standard deviation gives then a
measure of the number of monomers correlated to the perturbed monomer
$m(\tau)\simeq (2 \tau/\tau_0)^{1/2}$.  This diffusive scaling of the
tension propagation provides a basis for dynamics and rheology of Rouse
polymer in thermal environment.  For the active Rouse polymer, however,
the persistence time $\tau_A$ introduces the domain size $m_A \equiv
(2\tau_A /\tau_0)^{1/2}$~\cite{goy24}. Such a correlated domain is
built by the influence of the past persistent noise, indicating that
the original diffusive scaling applies to the transient case, but not
to the steady state due to the preexisting $m_A$. For the latter, the
number of dynamically correlated monomers exceeds $m_A$ only for $\tau
\gg \tau_A$, after which one recovers the standard tension propagation
scaling. We thus expect 
\begin{eqnarray} 
m(\tau) \simeq
    \left\{ \begin{array}{ccc}
       \left(\frac{2 \tau}{\tau_0} \right)^{1/2} &  ({\rm transient})\\
        m_A+\left(\frac{2 \tau}{\tau_0} \right)^{1/2}  &({\rm steady
        \ state})
    \end{array} \right.
\label{m_tau} 
\end{eqnarray} 
These scaling arguments are supported by the exact results of displacement
correlations from which one estimates the size of domains that perform
correlated motion \cite{long_paper}, see also below.

We combine now Eqs.~\eqref{msd-i} and~\eqref{m_tau}, as in the compounding 
formula~\eqref{CF2}, to obtain the following scaling predictions for the 
tagged monomer MSD for the AOU noise
\begin{itemize}
\item transient
\begin{eqnarray}
\langle \Delta z^2(n,\tau) \rangle_{tr}
\simeq \frac{A\tau_0^{1/2}}{\gamma^2}
\left\{
\begin{array}{ll} 
   \tau^{3/2}
   &  ( \tau \ll \tau_A) \\ 
  \tau_A \tau^{1/2}
  &  ( \tau \gg \tau_A)
\end{array}
\right.
\label{MSD_CF_prediction_tr}
\end{eqnarray}
\item steady state
\begin{eqnarray}
\langle \Delta z^2(n,\tau) \rangle_{ss}
\simeq \frac{A\tau_0^{1/2}}{\gamma^2}
\left\{
\begin{array}{ll} 
  \tau_A^{-1/2} \tau^2  &  ( \tau \ll \tau_A) \\ 
  \tau_A \tau^{1/2}  &  ( \tau \gg \tau_A)
\end{array}
\right.
\label{MSD_CF_prediction_ss}
\end{eqnarray}
\end{itemize}

A number of remarks on the compounding formula, in particular, its 
relevance to active polymers are in order.

(I) Earlier works predicted a MSD $\sim \tau^\alpha$ with
either $\alpha=3/2$~\cite{vand15,vand17,Sakaue_2017,Put_2019} or
$\alpha=2$~\cite{osma17,Gompper_2020} on short time scale.  A closer
inspection of these works shows that two different protocols were
used in the MSD calculation, with those in the former group using a
transient protocol, while the latter group using a steady-state one.
The results are in agreement with the compounding formula predictions
\eqref{MSD_CF_prediction_tr} and \eqref{MSD_CF_prediction_ss}.

(II) Some care is needed in dealing with the transient scaling
of Eq.~\eqref{m_tau}.  The tension propagation front $m(\tau) \sim
\tau^{1/2}$ only sets in on time scale $\tau > \tau_0 = \gamma/k$, the
monomer time scale. On shorter time scale the tagged monomer behaves as
being disconnected from the rest of the chain (or formally $m=1$ for
$\tau < \tau_0$, see Fig.~\ref{fig01} - top left).  This also happens
in the equilibrium Rouse chain in which the anomalous MSD scaling $\sim
\tau^{1/2}$ sets in on time scale $\tau > \tau_0$.

(III) In contrast, in the steady-state case, no change of scaling behavior
takes place at $\tau_0$. This is because on time scale $\leq \tau_A$
the number of dynamically connected monomers remains constant $m(\tau)
= m_A$, see  Fig.~\ref{fig01} - top right.

(IV) Despite the larger exponent for the steady state MSD ($\sim
\tau^2$), as opposed to the transient case ($\sim \tau^{3/2}$), the
actual displacement is smaller in the former case, e.g. $\langle \Delta
z^2(n,\tau) \rangle_{tr} \geq \langle \Delta z^2(n,\tau) \rangle_{ss}$,
as it can be verified by comparing Eqs.~\eqref{MSD_CF_prediction_tr} and
\eqref{MSD_CF_prediction_ss}.  Again, such a feature can be understood
through Eq.~\eqref{m_tau}: each monomer in steady state must move
collectively with $m_A$ consecutive monomers from the beginning.  Hence,
the tagged monomer dynamics reflects the center-of-mass mode of the
domain consisting of $m_A$ monomers. This indicates that the ballistic
scaling in steady state is super-universal, i.e., independent not only
of the details of noise statistics but also of the chain conformation. We
will later discuss how the latter influences the transient dynamics.

(V) On long time scale $\tau \gg \tau_A$, the persistence of the
active noise becomes irrelevant. One can then employ the white noise
description with an effective temperature, hence, the classical
Rouse scaling $\alpha=1/2$ emerges in both transient and steady
state protocols. However, if the active noise has slowly decaying
power-law memory $g(u) \sim u^{-\alpha_0}$ with $\alpha_0 < 1$,
Eq.~\eqref{isolated_monomer} leads to the isolated monomer MSD $\langle
\Delta z_i^2 (\tau)\rangle \approx  \tau^{2-\alpha}$ for $\tau \gg
\tau_A$, hence, the late time scale dynamics of the tagged monomer is
characterized by a nontrivial exponent $\alpha=3/2 - \alpha_0$.

{\it{Exact calculations --}}
Transient and steady state MSDs for a tagged monomer can be computed
building up from the Gaussian propagator \eqref{Green_func}.
This approach is different than the standard Rouse mode analysis, as
it considers the limit of a very long polymer $N \gg 1$.  We find the
following expressions \cite{long_paper}
\begin{eqnarray}
    \langle \Delta z^2 (n, \tau) \rangle_{t r}&=& 
    \frac{A}{\gamma^2} \sqrt{\frac{\tau_0}{\pi}} \!\!
      \int_0^\tau \!\! du \ g(u) \left[ \sqrt{2\tau -u} - \sqrt{u} \right]
    \nonumber \\
\label{app:res_tr}
\end{eqnarray}
\begin{eqnarray}
 \langle \Delta z^2 (n, \tau)\rangle_{ss} &=&
    \frac{A}{\gamma^2}  \sqrt{\frac{\tau_0}{\pi}} 
    \int_0^{+\infty} \!\!\!\!\!\!\!
    du  \,\, g(u)  \left( \sqrt{\tau+u} - \right.
    \nonumber\\
   && \left. 2 \sqrt{u} + \sqrt{|\tau-u|} \right) 
    \label{app:res_ss}
\end{eqnarray}
which are valid for any arbitrary noise correlator $g(u)$, also
beyond the AOU process. We use these two exact expressions with
\eqref{isolated_monomer} and \eqref{m_tau} to test the compounding
formula \eqref{CF2}.

Figure~\ref{fig:compound} shows a plot of the exact transient MSD
\eqref{app:res_tr} for a tagged monomer (solid lines) and of the
corresponding ratio of \eqref{isolated_monomer} and \eqref{m_tau}.
Three different types of noises are analyzed: (a) the AOU noise with
persistence time $\tau_A$, (b) a combination of two AOU noises with
persistence times $\tau_A/10$ and $10 \tau_A$, and (c) a power-law
correlated noise. In all three cases we find excellent agreement
between the exact calculation (solid lines) and the right hand side of
Eq.~\eqref{CF2} (dashed lines). The thin dotted lines are the asymptotic
short and long time scale behaviors. All noises show a universal $\sim
\tau^{3/2}$ scaling on short time scale which crosses over to a regime
$\sim \tau^{1/2}$ for the exponentially correlated noises (a, b). A
different long time behavior is seen for the power-law correlated noise
(c), see the above remark (V).

Figure~\ref{fig:compound-ss} analyzes the compounding formula
for the steady state regime for the same three noises used in
Fig.~\ref{fig:compound}.  The solid lines plot Eq.~\eqref{app:res_ss},
while the dotted lines are the right hand side of the compounding
formula \eqref{CF2}.  The graphs show that Eq.~\eqref{CF2} is verified
throughout a broad range of time scales. A universal $\sim \tau^{2}$
scaling on short time scale is clearly seen up to the time scale set by
the noise persistence.

We also analyzed the behavior of the displacement correlations
\begin{eqnarray}
    H(n,\tau) = \langle \Delta z(m,\tau) \Delta z (m+n,\tau) \rangle
\end{eqnarray}
again distinguishing between transient and steady state protocol.
We note that $H(0,\tau)$ is the MSD, while for generic $n$ it provides
information about the correlated motion of monomers separated by 
a distance $n$ along the chain. The exact analytical calculations of
$H(n,\tau)$ reveal two distinctive behaviors \cite{long_paper},
which are those shown schematically in Fig.~\ref{fig01}.
In the thermal equilibrium/transient case a correlated front
builds up on time scale $\tau > \tau_0$: all monomers within a distance
$n \leq (\tau/\tau_0)^{1/2}$ from a tagged monomers perform 
correlated motion, following the standard tension-propagation 
scaling. A different result is found for the polymer in an active 
steady state. We find, on time scale $\tau \ll \tau_A$, the
following displacement correlator \cite{long_paper}
\begin{eqnarray}
    H(n, \tau) \simeq \frac{A}{\gamma^2} G(n,\tau_A) \   \tau^2
    \label{H_a_ss_1}
\end{eqnarray}
with $G(n,\tau_A)$ the Gaussian propagator \eqref{Green_func}.
This prefactor indicates that all monomers in the range $n \leq
(\tau_A/\tau_0)^{1/2}$ perform correlated motion already at early time
scales, responding as a correlated block to the tagged monomer dynamics.
The inset of Fig.~\ref{fig:compound-ss} shows plots of $H(n,\tau)$
vs. $\tau$ on a log-log scale for equilibrium (red symbols) and active
steady-state (green symbols) dynamics.  $H_{eq}(n,\tau)$ is exponentially
small for early time scale until the monomer at distance $n$ is included
in the correlated block of monomers \cite{long_paper}.

{\it $\beta$-Models} --
While we focused here on the ordinary active Rouse model \eqref{eq:Rouse},
the validity of the compounding formula \eqref{CF2} was also verified
for the so-called $\beta$-models \cite{long_paper}. These models extend
the Rouse model by introducing the effect of long range connectivity
along the chain \cite{doi88,Amitai13}. They depend on a generalized
exponent $\eta$, which in equilibrium leads to a MSD scaling as
$\sim (\tau/\tau_0)^{1-1/\eta}$, with the ordinary Rouse limit
given by $\eta=2$. The tunability of the exponent $\eta$ allows for
interpolation through different regimes including the crumpled globule
($\eta=5/3$)\cite{tamm15,Sakaue_2017}, a model for an entanglement-free
large scale chromatin organization~\cite{Grosberg_1993}.  Our analysis
of active $\beta$-models shows that the compounding formula works and
provides an intuitive explanation of the various $\eta$-dependent scaling
regimes \cite{long_paper}. We note that the short time scaling in steady
state provides an MSD which is $\eta$-independent with a super-universal
scaling of the type $\sim \tau^2$ \cite{long_paper}.

\begin{figure}[t]
	\centering
    \includegraphics[width=0.49\textwidth]{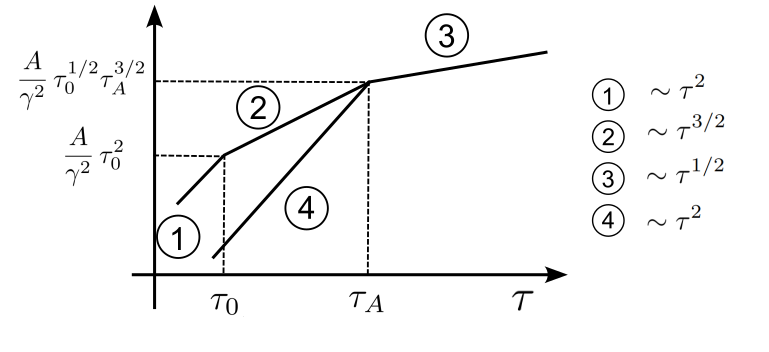}
	\caption{Summary of the scaling prediction of the tagged monomer
	MSDs $\langle \Delta z(n,\tau)^2 \rangle_{ss}$ and $\langle
	\Delta z(n,\tau)^2 \rangle_{tr}$ in active Rouse polymer kicked
	by persistent noise (AOU in this example).  Time evolution
	follows a sequence of regimes $\textcircled{\scriptsize
	1} \rightarrow \textcircled{\scriptsize 2} \rightarrow
	\textcircled{\scriptsize 3}$ in the transient protocol, and
	$\textcircled{\scriptsize 4} \rightarrow \textcircled{\scriptsize
	3}$ in the steady-state protocol. In \textcircled{\scriptsize
	3}, $\langle \Delta z(n,\tau)^2 \rangle_{ss} > \langle \Delta
	z(n,\tau)^2 \rangle_{tr}$ while sharing the same scaling
	exponent~\cite{long_paper}.}
	\label{fig_tra_ss}
\end{figure}

{\it Summary} --
  The compounding formula \eqref{CF2} is proven to be robust and to
  provide a solid theoretical framework through which one can analyze the
  dynamics of the active Rouse model. The main results are summarized
  in Fig.~\ref{fig_tra_ss}, which plots the various regimes of MSD for
  the steady state and transient dynamics.  We summarize the scaling
  behaviors here for the AOU noise.  For the transient protocol, 1: Up to
  a monomer time scale $\tau_0$ the monomer can be actually considered as
  free and under the effect of persistent noise the MSD scales as $\sim
  \tau^2$~\footnote{We note that this regime is not observed in the plots
  of Fig.~\ref{fig:compound} as the analytical solution \eqref{app:res_tr}
  is for a continuous model. In this $\tau \lesssim \tau_0$ regime, the
  monomer does not feel the connectivity with the rest of the chain and
  behaves as a free monomer.}.  2: In the time interval $\tau_0 \leq \tau
  \leq \tau_A$ the effect of chain connectivity becomes apparent. The
  tension propagation mechanisms sets up and the number of dynamically
  connected monomers grows as in Eq.~\eqref{m_tau}. This slows down the
  dynamics to an MSD scaling as $\sim \tau^{3/2}$. 3: Beyond $\tau \geq
  \tau_A$ the persistence of the active noise is lost and the scaling
  is as that of the thermal Rouse model $\sim \tau^{1/2}$.  The tension
  propagation mechanism is still active in this regime and the MSD scaling
  is due to the ratio of ordinary diffusive motion $\sim \tau$ divided
  by $\sim \tau^{1/2}$, which is the tension propagation contribution of
  increasing dynamically correlated monomers, Eq.~\eqref{CF2}. For the
  steady state protocol, 4:  the MSD scales as $\tau^2$ up to a time
  scale $\tau_A$. The steady active noise generates correlated active
  domains, which respond collectively to the persistent noise.

There are several directions to extend the present study. Among others,
the effect of inhomogeneous kicks, i.e., sparsity of the active noise
source would be interesting in the context of chromatin dynamics, see
refs~\cite{Jeon_2020,Jeon_2023,Goychuk23} for related works. In addition,
we point out possible relevance to other systems than polymers. One
example can be found in the problem of the growth of rough interfaces
\cite{Kardar_1986,Stanley_book,Takeuchi_2018}. It would be interesting
to examine the validity of the compounding formula, for instance, in
the growth model corresponding to the KPZ universality class with the
persistent noise.  Another system of interest is interacting particles
in narrow channels, i.e., single-file diffusion\cite{Ooshida_2016}. We
expect that our formalism applies to the dynamics of one-dimensional
array of active Brownian particles and also run-and-tumble particles
\cite{Chaudhuri_2024}.

{\sl  Acknowledgments} --
We wish to express our gratitude to the late Carlo Vanderzande who
introduced us to the field of active polymers.  T.S thanks J. Prost and
G.V. Shivashankar for useful discussions.  This work is supported by
JSPS KAKENHI (Grant No. JP23H00369 and JP24K00602).


%

\end{document}